\documentclass{wscpaperproc}
\usepackage{latexsym}
\usepackage{graphicx}
\usepackage{mathptmx}
\usepackage[T1]{fontenc}

\usepackage{amsmath}
\usepackage{amsfonts}
\usepackage{amssymb}
\usepackage{amsbsy}
\usepackage{amsthm}
\usepackage[pdftex,colorlinks=true,urlcolor=blue,citecolor=black,anchorcolor=black,linkcolor=black]{hyperref}
\usepackage{subcaption}
\usepackage{multirow}
\usepackage{xcolor}
\newtheoremstyle{wsc}% hnamei
{3pt}% hSpace abovei
{3pt}% hSpace belowi
{}% hBody fonti
{}% hIndent amounti1
{\bf}% hTheorem head fontbf
{}% hPunctuation after theorem headi
{.5em}% hSpace after theorem headi2
{}% hTheorem head spec (can be left empty, meaning `normal')i

\theoremstyle{wsc}

\begin{document}

%***************************************************************************
% AUTHOR: AUTHOR NAMES GO HERE
% FORMAT AUTHORS NAMES Like: Author1, Author2 and Author3 (last names)
%
%		You need to change the author listing below!
%               Please list ALL authors using last name only, separate by a comma except
%               for the last author, separate with "and"
%

% setting up general page style
\pagestyle{fancyplain}

% setting up page style of first page
\thispagestyle{plain}
\firstPageHead{}

% setting up running header (authors) of subsequent pages
\chead{\fancyplain{}{\itshape Dwarakanath, Vyetrenko, Oyebode, and Balch}}

% setting up seperation parameters
%\headsep=72pt
\rhead{}
\cfoot{}
\renewcommand{\headrulewidth}{0pt} % (renewcommand needed in fancyhdr to remove top decorative line)
%\headrulewidth=0pt  % ("setlength" needed in fancyheading to remove top decorative line)

%%%%%%%%%%%%%%%%%%%%%%%%%%%%%%%%%%%%%%%%%%%%%%%%%%%%%%%%%%%%%%%%%%%%%%%%%%%%%%
%                                                                            %
%     THESE COMMANDS ARE REQUIRED TO WORK WITH WSC.BST TO MAKE BIBLIO     %
%                                                                            %
%%%%%%%%%%%%%%%%%%%%%%%%%%%%%%%%%%%%%%%%%%%%%%%%%%%%%%%%%%%%%%%%%%%%%%%%%%%%%%
\makeatletter
\let\@internalcite\cite
\def\cite{\def\@citeseppen{-1000}%
    \def\@cite##1##2{(##1\if@tempswa , ##2\fi)}%
    \def\citeauthoryear##1##2##3{##1 ##3}\@internalcite}
\def\citeNP{\def\@citeseppen{-1000}%
    \def\@cite##1##2{##1\if@tempswa , ##2\fi}%
    \def\citeauthoryear##1##2##3{##1 ##3}\@internalcite}
\def\citeN{\def\@citeseppen{-1000}%
%  Pierre L'Ecuyer's fix for multiple cite bug
%  Added by Paul J Sanchez on 4 October 2001
%   \def\@cite##1##2{##1\if@tempswa , ##2)\else{)}\fi}%
%   \def\citeauthoryear##1##2##3{##1 (##3}\@citedata}
    \def\@cite##1##2{##1\if@tempswa, ##2)\else{}\fi}%
    \def\citeauthoryear##1##2##3{##1 (##3)}\@citedata}
\def\citeA{\def\@citeseppen{-1000}%
    \def\@cite##1##2{(##1\if@tempswa , ##2\fi)}%
    \def\citeauthoryear##1##2##3{##1}\@internalcite}
\def\citeANP{\def\@citeseppen{-1000}%
    \def\@cite##1##2{##1\if@tempswa , ##2\fi}%
    \def\citeauthoryear##1##2##3{##1}\@internalcite}
\def\shortcite{\def\@citeseppen{-1000}%
    \def\@cite##1##2{(##1\if@tempswa , ##2\fi)}%
    \def\citeauthoryear##1##2##3{##2 ##3}\@internalcite}
\def\shortciteNP{\def\@citeseppen{-1000}%
    \def\@cite##1##2{##1\if@tempswa , ##2\fi}%
    \def\citeauthoryear##1##2##3{##2 ##3}\@internalcite}
\def\shortciteN{\def\@citeseppen{-1000}%
%  Pierre L'Ecuyer's fix for multiple cite bug
%  Added by Paul J Sanchez on 2 September 2002
%  should have caught this last year...
%   \def\@cite##1##2{##1\if@tempswa , ##2)\else{)}\fi}%
%   \def\citeauthoryear##1##2##3{##2 (##3}\@citedata}
% Shane G. Henderson fix for extra right bracket at end of optional material June 8, 2005
%    \def\@cite##1##2{##1\if@tempswa, ##2)\else{}\fi}%
    \def\@cite##1##2{##1\if@tempswa, ##2\else{}\fi}%
    \def\citeauthoryear##1##2##3{##2 (##3)}\@citedata}
\def\shortciteA{\def\@citeseppen{-1000}%
    \def\@cite##1##2{(##1\if@tempswa , ##2\fi)}%
    \def\citeauthoryear##1##2##3{##2}\@internalcite}
\def\shortciteANP{\def\@citeseppen{-1000}%
    \def\@cite##1##2{##1\if@tempswa , ##2\fi}%
    \def\citeauthoryear##1##2##3{##2}\@internalcite}
\def\citeyear{\def\@citeseppen{-1000}%
    \def\@cite##1##2{(##1\if@tempswa , ##2\fi)}%
    \def\citeauthoryear##1##2##3{##3}\@citedata}
\def\citeyearNP{\def\@citeseppen{-1000}%
    \def\@cite##1##2{##1\if@tempswa , ##2\fi}%
    \def\citeauthoryear##1##2##3{##3}\@citedata}
%
% \@citedata and \@citedatax:
%
% Place commas in-between citations in the same \citeyear, \citeyearNP,
% \citeN, or \shortciteN command.
% Use something like \citeN{ref1,ref2,ref3} and \citeN{ref4} for a list.
%
\def\@citedata{%
    \@ifnextchar [{\@tempswatrue\@citedatax}%
                  {\@tempswafalse\@citedatax[]}%
}

\def\@citedatax[#1]#2{%
\if@filesw\immediate\write\@auxout{\string\citation{#2}}\fi%
  \def\@citea{}\@cite{\@for\@citeb:=#2\do%
    {\@citea\def\@citea{, }\@ifundefined% by Young
       {b@\@citeb}{{\bf ?}%
       \@warning{Citation `\@citeb' on page \thepage \space undefined}}%
{\csname b@\@citeb\endcsname}}}{#1}}%

% don't box citations, separate with ; and a space
% also, make the penalty between citations negative: a good place to break.
%
\def\@citex[#1]#2{%
\if@filesw\immediate\write\@auxout{\string\citation{#2}}\fi%
  \def\@citea{}\@cite{\@for\@citeb:=#2\do%
    {\@citea\def\@citea{; }\@ifundefined% by Young
       {b@\@citeb}{{\bf ?}%
       \@warning{Citation `\@citeb' on page \thepage \space undefined}}%
{\csname b@\@citeb\endcsname}}}{#1}}%

% (from apalike.sty)
% No labels in the bibliography.
%
\def\@biblabel#1{}
\makeatother

%\newlength{\bibhang}
%\setlength{\bibhang}{2em}

% Indent second and subsequent lines of bibliographic entries. Taken
% from openbib.sty: \newblock is set to {}.
% \renewcommand{\refname}{REFERENCES}

\newdimen\bibindent
\bibindent=0.0em
% SEC: was \def\thebibliography#1{\section*{\refname\@mkboth
% SEC: was   {\uppercase{\refname}}{\uppercase{\refname}}}\list
\def\thebibliography#1{\section*{\refname}\list
   {}{\settowidth\labelwidth{[#1]}
   \leftmargin\parindent
   \itemindent -\parindent
   \listparindent \itemindent
   \itemsep 0pt
   \parsep 0pt}
   \def\newblock{}
   \sloppy
   \sfcode`\.=1000\relax}

           % Set up BiBTeX macros

% needed to make the tex document look more like the word counterpart :-(
\setlength{\baselineskip}{12.7pt}

% AUTHOR: Enter the title, all letters in upper case
\title{Transparency as Delayed Observability in Multi-Agent Systems}

% AUTHOR: Enter the authors of the siarticle, see end of the example document for further examples
\author{Kshama Dwarakanath\\
Svitlana Vyetrenko\\
Tucker Balch\\[12pt]
	JP Morgan AI Research\\
 383 Madison Avenue\\
 New York, NY 10179, USA\\
% Multiple authors are entered as follows.
% You may also need to adjust the titlevbox size in the preamble - search for titlevboxsize
\and
Toks Oyebode\\ \\ \\[12pt]
    JP Morgan Regulatory Affairs\\
    25 Bank Street, Canary Wharf\\
	London, E14 5JP, UK\\
}
% \author{Kshama Dwarakanath\\[12pt]
% 	JP Morgan AI Research\\
% 	Palo Alto, CA, USA\\
% % Multiple authors are entered as follows.
% % You may also need to adjust the titlevbox size in the preamble - search for titlevboxsize
% \and
% Svitlana Vyetrenko\\[12pt]
%     JP Morgan AI Research\\
% 	Palo Alto, CA, USA\\
% \and
% Toks Oyebode\\ [12pt]
%     JP Morgan Regulatory Affairs\\
% 	London, UK\\
% \and
% Tucker Balch\\ [12pt]
%     JP Morgan AI Research\\
% 	New York, NY, USA\\
% }

\maketitle

\section*{ABSTRACT}
Is transparency always beneficial in complex systems such as traffic networks and stock markets? How is transparency defined in multi-agent systems, and what is its optimal degree at which social welfare is highest? 
We take an agent-based view to define transparency (or its lacking) as delay in agent observability of environment states, and utilize simulations to analyze the impact of delay on social welfare. 
To model the adaptation of agent strategies with varying delays, we model agents as learners maximizing the same objectives under different delays in a simulated environment. 
Focusing on two agent types - constrained and unconstrained, we use multi-agent reinforcement learning to evaluate the impact of delay on agent outcomes and social welfare.
Empirical demonstration of our framework in simulated financial markets shows opposing trends in outcomes of the constrained and unconstrained agents with delay, with an optimal partial transparency regime at which social welfare is maximal. 

\section{INTRODUCTION}
Multi-agent systems (MASs) are ubiquitous in applications such as robotics \shortcite{dudek1996taxonomy}, healthcare \cite{shakshuki2015multi}, finance \shortcite{byrd2020abides} and transportation \shortcite{burmeister1997application} where the domain can be modeled as a system with interacting agents/components. Each agent is an entity that takes in sensory observations of the environment to make goal-oriented decisions \shortcite{dorri2018multi}. There are numerous challenges associated with the design and development of MASs including communication and collaboration between agents, multi-agent learning and lastly, transparency or the degree of information dissemination. Since each agent in a MAS uses disseminated information to update its decisions, modifying transparency can lead to contrasting agent and system behaviors. Concurrently, simulations offer an effective framework to understand the impact of transparency on MASs to subsequently inform real world policy.

In this work, we consider stochastic MASs comprised of adaptive agents. The stochastic nature arises from various factors including intrinsic environmental randomness (uncontrollable by agents), randomness in agent behavior, along with the fact that the system evolves as a result of agent interactions. Each agent has access to a set of system observables that potentially informs their behavior. The more informative the set of observables are, the more knowledge the agent has about the system including the behavior of other agents. \textbf{We characterize transparency in such stochastic MASs by the degree of observability of agents.} A system comprised of agents with a higher degree of observability is said to be more transparent than one where agents have a lower degree of observability. 

We seek to analyze the impact of varying observability in stochastic MASs on the strategies adopted by the agents, and subsequently on social welfare. We mathematically formulate a type of observability called delayed observability that is characterized by a single delay parameter controlling the degree of observability. 
% We propose a mathematical formulation of a specific type of observability, called delayed observability. Systems with delayed observability are characterized by a single delay parameter that controls the degree of observability. 
We then let the agents adapt their strategies to this delay parameter. In order not to hard code (and bias/restrict) this adaptation, we equip agents with learning algorithms designed to maximize their objectives under varying delays. 
% This prevents the introduction of any bias into the change in agent strategies with observability. 
Therefore, we formulate this as a multi-agent reinforcement learning problem with partial observability. The key contributions of this work are as follows.\begin{enumerate}
    \item A mathematical formulation for transparency in multi-agent systems using the notion of delayed observability in stochastic games.
    \item A framework for analyzing the impact of observability on strategies and outcomes of constrained and unconstrained agents, and on social welfare using multi-agent reinforcement learning. We adapt a social welfare metric that captures agents' average outcome as well as equality of outcomes.
    \item Detailed empirical study of the proposed framework in simulated financial markets. We use a multi-agent market simulator to train our agents, and investigate the variation in their policies with observability. Our findings indicate the trade-offs between the degree of observability and agent outcomes, and suggest an intermediate degree of observability at which social welfare is maximized. 
\end{enumerate}

\section{BACKGROUND AND RELATED WORK}

\subsection{Transparency}

Transparency has been of interest in various disciplines including governance, medicine, financial markets and organisations \cite{ball2009transparency}. It has diverse meanings ranging from being a tool to affect accountability and performance of governmental agencies \cite{kosack2014does} to one that builds patient trust in medical physicians \shortcite{chimonas2017bringing}, and stakeholder trust in organisations \cite{auger2014trust}. Regulators of financial markets are increasingly focused on improving information dissemination to traders to improve market efficiency \cite{sec_bonds,rfi}. Recent work \shortcite{barsotti2022transparency} looks at the interplay between transparent explanations shared by an institution and strategic adaptation by individuals subject to a classification model. They evaluate the impact of such feedback on faking behaviour by individuals.
%and detection capacity of the institution.

% Examples of transparency
% \begin{itemize}
%     \item Transparency in financial markets 
%     % \item Transparency in governance \cite{kosack2014does} -- transparency as tool to affect accountability and performance of the government
%     % \item Transparency in medicine \cite{chimonas2017bringing} -- transparency to build patient trust in physicians
%     % \item Transparency in organisations \cite{auger2014trust} -- impact of transparent communication on trust and intentions of stakeholders in organizations
% \\
%     \cite{jehiel2015transparency} -- optimal distribution of information in organisations so as to align goals of agents with those of the organization
%     % \item General def of transparency \cite{ball2009transparency} -- classifying transparency and how their meanings affect organisations including the governmental ones
% \end{itemize}

We look at transparency as information availability to agents in a dynamic MAS that can be used to improve their objectives. There is work on studying the impact of information in networks of interacting agents in the game theory literature that involves classifying information in games into two types \shortcite{comp_obser}. The first type looks at the availability of the rules of the game to players in it, resulting in games with complete or incomplete information \cite{harsanyi}. The second type examines the visibility of agent actions to one another thus partitioning games into those with perfect or imperfect information.  

\shortciteN{ARNOTT1991309} challenge the intuitive notion that increased information dissemination in traffic systems reduces congestion by investigating the equilibrium effects of such information on driver behaviors as well as mean travel times. Through mathematical models for commuter travel times and costs, the authors make the case that while proprietary information may benefit a single driver, informing all drivers can result in them being worse off than when uninformed. \shortciteN{das2017reducing} look at designing information made available to agents in a congestion game in order to improve social welfare. The authors make the case for partial information being able to improve efficiency in such network routing games. 

While transparency in systems is generally a positive notion as it is intuitively thought to improve information access and opportunity for participants, there exists literature in the financial domain to investigate this intuition deeper. They account for the fact that individual agent objectives in MASs may not be in line with each other or with overall social welfare, causing the MAS to be more sensitive to released information. \citeN{purdah} look into existing practice by central banks to refrain from information transparency right before policy meetings because transparency during these periods could lead to higher market volatility. The authors corroborate this rationale by analyzing past market data 
% and fitting time series models to asset prices 
to empirically estimate effects of information dissemination. \citeN{walsh2007optimal} and \shortciteN{van2010optimal} explore the optimal degree of central bank transparency that can limit inflation. Here, they measure transparency by central bank announcements about its view on the economy or its own private information. 

\subsection{Partially Observable Stochastic Games and Multi-Agent Reinforcement Learning\label{subsec:posg}}
We seek to evaluate how strategic agents in a MAS utilize information available under increased transparency regimes to adapt their behaviours. Given agent objectives, we allow these agents to use information to learn strategies that improve their objectives over time. A single agent that is learning to act in an uncertain environment is modeled by a Markov Decision Process (MDP) in the framework of reinforcement learning \cite{sutton2018reinforcement}. In the MDP framework, multiple agents are modeled to be non-adaptive and accounted as part of the environment of the learning agent. Multiple adaptive agents with interacting or competing goals can be modeled by Markov Games in multi-agent reinforcement learning. Markov Games are an extension of MDPs involving the specification of a global state space, action spaces for each agent along with corresponding reward functions. \citeN{littman1994markov} propose a Q-learning algorithm for 2-player zero-sum games where the objective of the first player is exactly opposite to that of the second. 

An MDP is said to be partially observable (PO) if the environment state is not completely visible to the agent \cite{spaan2012partially}. Partially observable MDPs can be extended to PO Markov Games (or PO Stochastic Games) to precisely model stochastic MASs with varying degrees of observability \shortcite{shapley1953stochastic,hansen2004dynamic}. A finite horizon Partially Observable Stochastic Game (POSG) is denoted by $\Gamma=\langle \mathcal{N},\mathcal{S},\lbrace\mathcal{A}_i\rbrace_{i=1}^n,\lbrace\mathcal{O}_i\rbrace_{i=1}^n,\mathbb{T},\lbrace \mathbb{O}_i\rbrace_{i=1}^n,\lbrace R_i\rbrace_{i=1}^n,\gamma,H\rangle$ where \begin{itemize}
    \item $\mathcal{N}=\lbrace1,2,\cdots,n\rbrace$ is the set of agents
    \item $\mathcal{S}$ is the state space
    \item $\mathcal{A}_i$ is the action space of agent $i$ with $\mathcal{A}=\mathcal{A}_1\times\mathcal{A}_2\times\cdots\times\mathcal{A}_n$ denoting the joint action space
    \item $\mathcal{O}_i$ is the observation space of agent $i$ %for $i\in\mathcal{N}$ 
    \item $\mathbb{T}:\mathcal{S}\times\mathcal{A}\rightarrow\mathbb{P}\left(\mathcal{S}\right)$ is the transition function that maps the current state and joint action to a probability distribution over the next state
    \item $\mathbb{O}_i:\mathcal{S}\rightarrow\mathbb{P}\left(\mathcal{O}_i\right)$ is the observation function that maps the current state to a probability distribution over observations of agent $i$ %for $i\in\mathcal{N}$
    \item $R_i:\mathcal{S}\times\mathcal{A}\rightarrow\mathbb{R}$ is the reward function of agent $i$ %for $i\in\mathcal{N}$
    \item $\gamma\in[0,1)$ is a discount factor
    \item $H$ is the horizon
\end{itemize}
The objective of each agent $i\in\mathcal{N}$ in a POSG is to find a sequence of their own actions that maximizes their expected sum of discounted rewards over the horizon\begin{align}
    \max_{\left(a_i(0),\cdots,a_i(H-1)\right)}\ \mathbb{E}\left[\sum_{t=0}^{H-1}\gamma^tR_i\left(s(t),a_1(t),\cdots,a_n(t)\right)\right]\nonumber
\end{align}
where $s(t+1)\sim\mathbb{T}\left(s(t),a_1(t),\cdots,a_n(t)\right)\ \forall t$. A POSG is fully observable when $\mathcal{O}_i=\mathcal{S}$ with $\mathbb{O}_i$ being the Dirac delta function over $\mathcal{O}_i=\mathcal{S}$ for all $i\in\mathcal{N}$. 
% \cite{dirac_delta_wiki}.

\shortciteN{comp_obser} take a partial observability approach to optimal information transparency in network games where a principal needs to decide on agent neighborhoods for aggregate information sharing. They formulate this as an optimization problem seeking an optimal partition of agents to maximize social welfare. \citeN{dist_submodular} look at the effect of a lack of information about other agents' strategies on agent objectives. These two works differ from ours in that transparency refers to the knowledge of other agents' actions, rather than an observation of a global system state as in our case. We also deal with applications where the change in agent strategies in response to those of others is not easily defined due to the dynamic nature of their objectives, calling for the use of multi-agent simulations. Thus, we allow agents to adapt their strategies in order to maximize their objectives using reinforcement learning. 

\section{PROBLEM FORMULATION}
% We learn agent strategies that optimize their objectives for each observability regime, and compare them across regimes to understand the impact of varying observability. 

\subsection{Delayed Observability and Constrained Agents}
In this work, we define the lack of transparency in MASs as the delay in observability in their POSG formulation. We introduce the notion of \textbf{delayed observability} as follows. Partition every state $s\in\mathcal{S}$ as $s=\begin{bmatrix}
    s_{I}&s_{D}
\end{bmatrix}$ where $s_I$ is a part of $s$ that is immediately observable to all agents, while $s_D$ is a part of $s$ that is observable after a time delay $\delta\in[0,H]$. Thus, the observation of any agent $i\in\mathcal{N}$ at time $t$ is \begin{align}
    o_i(t)&=\begin{bmatrix}
        s_I(t)&s_D(t-\delta)
    \end{bmatrix}\label{eq:delayed_obs}
\end{align} 
Note that we do not consider $\delta>H$ since that would correspond to looking into states from previous episodes. Equivalently, we formulate POSGs with delayed observability $\delta$ as those where $\mathcal{O}_i=\mathcal{S}$ and the observation function $\mathbb{O}_i$ is a indicator for observations satisfying (\ref{eq:delayed_obs}), for all agents $i\in\mathcal{N}$. Note that a large delay $\delta$ would correspond to less transparency, and vice-versa. Additionally, we do not model any hidden and fully unobservable states in this work thereby eliminating the need for belief state updates.

For an MDP with partial observability (i.e. a POSG with a single agent), \citeN{aastrom1965optimal} showed that the agent's cumulative rewards increase with increase in observability. In this work, we look at the impact of the delay in observability $\delta$ on the strategies learnt by 2 players/agents in the POSG. These players are characterized by having the same reward function, but the actions of player 1 are \textit{more constrained} than those of player 2. More rigorously, we have $\mathcal{A}_1\subsetneq\mathcal{A}_2$ with $R_1(s,a_1,a_2)=R_2(s,a_1,a_2)\forall s\in\mathcal{S},a_i\in\mathcal{A}_i$ (and $\mathcal{O}_1=\mathcal{O}_2$ from before). We call player 1 the \textbf{constrained player} and player 2 the \textbf{unconstrained player}. 

In financial markets, constrained players include market makers whose actions are subject to regulatory constraints of liquidity provision \cite{chakraborty2011market}. And, unconstrained players would be firms trading similar volumes as market makers, without being constrained to provide liquidity. In traffic systems, constrained players would be (public) government transit systems that are restricted to a smaller set of routes as opposed to other (unconstrained) commuters or private transit agencies. It is then natural to expect that increased observability affects unconstrained players in a different fashion than constrained players. The interaction between the two leading to potentially interesting implications to social welfare. 

\subsection{Social Welfare Function}
% {\color{blue}provide more clarity (maybe in subsections) on the two versions of the social welfare function - maybe subsection (a) and (b) - and then follow it through to the chart in a later section}
Social welfare refers to the notion of goodness of current state of affairs with respect to the society as a whole \cite{sen2018collective}. It can be quantified through social welfare functions (SWFs) that can rank social states as being less or more desirable for social welfare. Given utilities/outcomes for individuals in a population, there are many SWFs studied in the literature on optimal taxation. 
% For example, the utilitarian social welfare function uses the sum of individual outcomes as a metric for social welfare. 
Let $Y_i\left(s(0),\pi_1(\cdot),\cdots,\pi_n(\cdot)\right)$ denote the outcome for player $i$ over horizon $H$ when players use respective policies $\pi_j:\mathcal{S}\rightarrow\mathcal{A}_j,\ \forall j\in\mathcal{N}$ starting from game state $s(0)$. For player outcomes $Y=\begin{bmatrix}Y_1&\cdots&Y_n\end{bmatrix}$, the utilitarian SWF is $\textnormal{SWF}(Y)=\sum_{i\in\mathcal{N}}Y_i$. 

Given an average population outcome $\Bar{Y}=\frac{1}{n}\sum_{i\in\mathcal{N}}Y_i$, different outcome distributions between players correspond to different levels of equality. We draw from \shortciteN{zheng2022ai} in capturing the trade-off between profitability and equality by using a product of both as a SWF as\begin{align}
    \textnormal{SWF}(Y)=\textnormal{Equality}(Y)\times \textnormal{Profitability}(Y)\nonumber
\end{align}
with profitability measured by $\Bar{Y}$. Popular (in)equality metrics in the social choice literature include the Gini index \cite{sen2018collective} and the Theil-L index \shortcite{sen1997economic}. Since we have two types of players (likely with multiple players of each type), we use generalized entropy indices to measure (in)equality as in \shortciteN{dwarakanath2022equitable,speicher2018unified}. They are attractive due to their property of subgroup decomposability while containing several inequality indices as special cases. 

We consider the following SWFs (although our framework is flexible to the use of any other):
\begin{enumerate}
\item Using the generalized entropy index with parameter $\kappa\notin\lbrace0,1\rbrace$ to measure equality as\begin{align}
    \textnormal{SWF}(Y)=\exp{\left(-\textnormal{GE}_{\kappa}(Y)\right)}\times\Bar{Y}\label{eq:swf1}
    \end{align}
    where $\textnormal{GE}_{\kappa}(Y)=\frac{1}{n\kappa(\kappa-1)}\sum_{i\in\mathcal{N}}\left[\left(\frac{Y_i}{\Bar{Y}}\right)^\kappa-1\right]$.
% \begin{align}
%     \textnormal{GE}_{\kappa}(Y)=\frac{1}{n\kappa(\kappa-1)}\sum_{i\in\mathcal{N}}\left[\left(\frac{Y_i}{\Bar{Y}}\right)^\kappa-1\right].\nonumber
% \end{align}
\item Using the Theil-L index given by $\textnormal{Theil}_{L}(Y)=-\frac{1}{n}\sum_{i\in\mathcal{N}}\ln\left(\frac{Y_i}{\Bar{Y}}\right)$ to measure equality as
\begin{align}
    \textnormal{SWF}(Y)=\exp{\left(-\textnormal{Theil}_{L}(Y)\right)}\times\Bar{Y}.\label{eq:swf2}
    \end{align}
    % where the Theil-L index is given by $\textnormal{Theil}_{L}(Y)=-\frac{1}{n}\sum_{i\in\mathcal{N}}\ln\left(\frac{Y_i}{\Bar{Y}}\right)$.
%     \begin{align}
%     \textnormal{Theil}_{L}(Y)=-\frac{1}{n}\sum_{i\in\mathcal{N}}\ln\left(\frac{Y_i}{\Bar{Y}}\right)\nonumber
% \end{align}
\end{enumerate}
The negative sign in the exponential in (\ref{eq:swf1})-(\ref{eq:swf2}) is because these indices are metrics for inequality, with the generalized entropy index containing the Theil-L index as a special case when $\kappa=0$.
% \begin{align}
%     \textnormal{GE}_{\kappa}(Y)=\begin{cases}\frac{1}{n\kappa(\kappa-1)}\sum_{i\in\mathcal{N}}\left[\left(\frac{Y_i}{\Bar{Y}}\right)^\kappa-1\right],\textnormal{ for }\kappa\notin\lbrace0,1\rbrace\\
%     -\frac{1}{n}\sum_{i\in\mathcal{N}}\ln\left(\frac{Y_i}{\Bar{Y}}\right),\textnormal{ if }\kappa=0
%     \end{cases}\nonumber
% \end{align}
% where $\kappa=0$ corresponds to the Theil-L index. 
% Although we adopt (\ref{eq:swf1})-(\ref{eq:swf2}) as the social welfare functions in this work, our framework is flexible to the use of any other. 

The goal of this work is to evaluate the impact of the delay in observability $\delta$ on 
\begin{itemize}
    \item strategies learnt by the (constrained) player 1 and (unconstrained) player 2 
    \item outcomes of both players resulting from playing out learnt strategies
    \item social welfare measured by (\ref{eq:swf1}) and (\ref{eq:swf2})
\end{itemize}
We emphasize that we equip both players with learning algorithms to learn strategies that utilize available observations to maximize their objectives without any constraints on the way strategies change with delay in observability. Thus, we formulate this problem as a 2-agent reinforcement learning problem for every value of $\delta$. We specifically experiment with the example of simulated financial markets. 

\section{APPLICATION TO MARKETS}
% We now describe the multi-agent financial market simulation framework along with its POSG formulation with delayed observability. 

\subsection{Multi-Agent Market Simulator}

\begin{figure}[h!]
    \centering
    \includegraphics[width=0.5\linewidth]{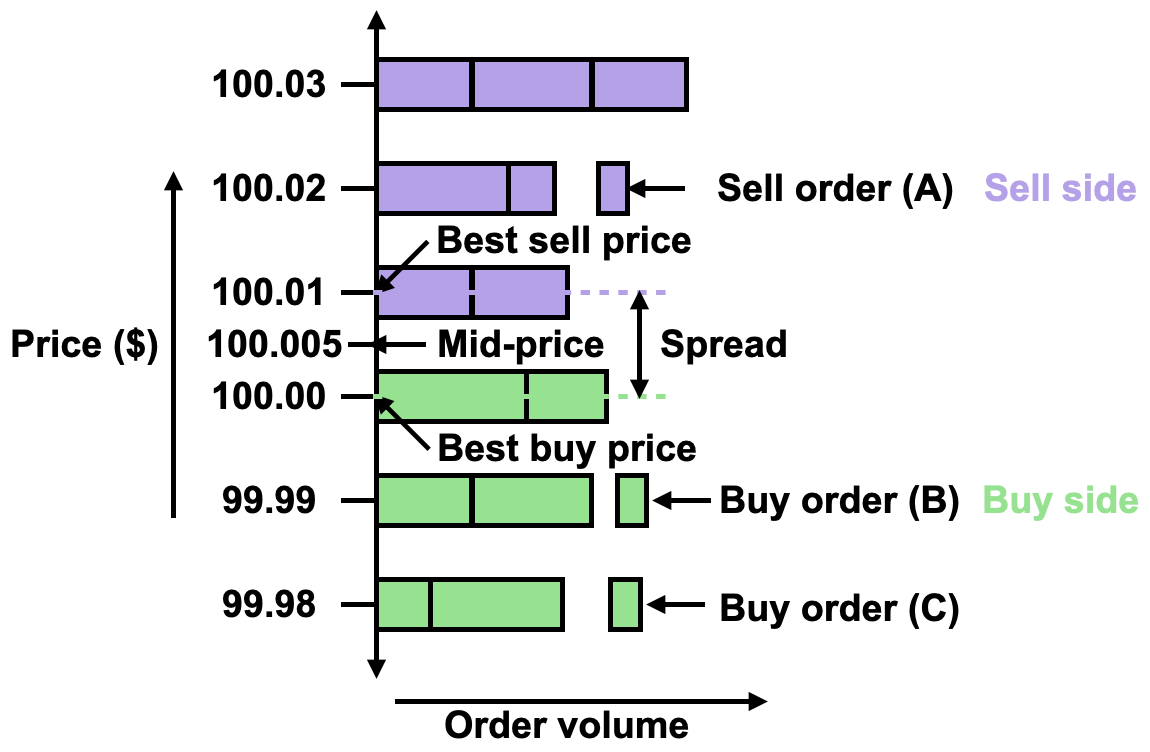}
    \caption{Snapshot of buy and sell orders at an exchange. Mid-price moves when traders submit orders that cross the spread i.e., a buy order with price greater than or equal to the best sell price or vice-versa.}
    \label{fig:lob}
\end{figure}

Consider exchange-based markets (e.g. US equities market) comprising a variety of traders that send their order requests to a centralized exchange that matches buy and sell orders. Traders are allowed to specify both the price and the direction (buy or sell) of their orders sent to the exchange.
% wherein an exchange handles interactions between traders of different types. 
In order to simulate trades in such a market, we employ a multi-agent market simulator called ABIDES \shortcite{byrd2020abides,amrouni2021abides}. ABIDES provides a selection of trading agents with different trading incentives and behaviors. The simulation engine handles all communication between trading agents and the exchange. Figure \ref{fig:lob} shows an example snapshot of orders collected at the exchange wherein each purple/green rectangle represents a block of sell/buy orders respectively. The price of the cheapest sell order is called the best sell price while that of the most expensive buy order is called the best buy price. The current stock price (also called mid-price) is the average between the best sell and best buy prices, with the difference between the best sell and best buy prices called the spread. 
Figure \ref{fig:lob} shows an example with a mid-price of $10000.5$ cents and spread of $1$ cent.  
We populate the simulated market with the following traders:\begin{enumerate}
    \item \textbf{Market Maker (MM)}: Player 1 that is constrained to place orders on both the buy and sell sides of the market, following from its regulatory definition \shortcite{wah2017welfare,frb_dealers}. MM can choose price levels for the buy, sell orders relative to the current stock price, and earns the difference between the sell and buy prices if orders are matched on both sides. 
    %In this work, we define a MM by the stylized parameters that follow from its regulatory definition \cite{wah2017welfare,chakraborty2011market}. 
    % At every time $t$, the MM places new price quotes of constant order size $I$ at $d$ price increments around the current stock price $p_t$ in cents i.e., it places buy orders at prices $p_t-h-d, \ldots, p_t-h$ and sell orders at prices $p_t + h, \ldots, p_t + h +d$, where $d$ is the depth of placement and $h$ is the half-spread chosen by the MM at time $t$. 
    \item \textbf{Principal Trader (PT)}: Player 2 that can choose to place either a buy order, a sell order, buy and sell orders, or hold (no order), along with the price at which to do so relative to the current stock price. PT is hence unconstrained, with the volumes of each buy/sell order being equal for both MM and PT. 
    \item \textbf{Background traders}: Non-adaptive traders that utilize available market signals to trade in a rule based fashion. They include consumer traders employing a uniformly random trading strategy with random arrivals to the market, as well as intelligent traders that use momentum strategies or exogenous stock value to trade \shortcite{byrd2020abides}.  
    % Their strategies for trading include rnadom comparing
    % They include traders of three types: fundamental, momentum and consumer traders. The fundamental traders trade in line with their belief of the exogenous stock value (which we call fundamental price), without any view of the market microstructure \cite{kyle1985continuous}. The momentum traders follow a simple momentum strategy of comparing a long-term average of the price with a short-term average. If the short-term average is higher than the long-term average, the trader buys since the price is seen to be rising. And, vice-versa for selling. Consumer traders trade purely based on demand, coming in once a day to place an order of a random size in a random direction (buy or sell). \textcolor{red}{This much explanation necessary?}
\end{enumerate}

\subsection{POSG Formulation for Markets \label{subsec:posg_form_markets}}
The \textbf{state} of the environment at time $t$ includes \begin{itemize}
    \item \textit{Quotes}: Prices and volumes of quoted orders (pre-trading) on buy, sell sides over period $t-L,\cdots,t$
    \item \textit{Spread}: Market spread defined as the difference between the best sell and best buy prices at $t$
    \item \textit{Depth}: Half of the price difference between worst sell (most expensive sell) and worst buy (cheapest buy) orders at $t$
    \item \textit{Inventory}: Volume of stocks held by players 1 and 2 over $t-1,t$
    \item \textit{Cash}: Cash held by players 1 and 2 over $t-1,t$
    % \item \textit{Executed volume}: Volume of MM orders executed between $t-1$ and $t$ on buy and sell sides
    \item \textit{Momentum}: Momentum signals for stock price over 1, 10 and 30 time steps defined as the ratio of current mid-price to that 1/10/30 time steps before
    \item \textit{Trades}: Price and volumes of traded buy and sell orders over period $t-M,\cdots,t$
\end{itemize}
Note that the length of history for quotes $L$ and that for trades $M$ are hyper-parameters. 

The \textbf{observation} for player $i\in\mathcal{N}$ at time $t$ includes the immediately observable states of \textit{Quotes}, \textit{Spread}, \textit{Depth}, \textit{Inventory} of player $i$, \textit{Cash} of player $i$ and \textit{Momentum}. It also includes delayed \textit{Trades} information with delay parameter $\delta$. Hence at $t$, player $i$ has access to traded volumes and prices over the period $t-M-\delta,\cdots,t-\delta$. Intuitively, low $\delta$ implies knowledge of more recent trades in the market and hence, access to more relevant information than for high $\delta$.

The \textbf{action} for (constrained) player 1 represented by the MM includes\begin{itemize}
    \item \textit{Half-spread}: Distance from current stock price at which MM symmetrically places orders on the buy and sell sides. Figure \ref{fig:lob} shows an example snapshot of MM orders (A) and (B) placed at the exchange with half-spread $1.5$ cents when the stock price is $10000.5$ cents. 
\end{itemize}

The \textbf{action} for (unconstrained) player 2 represented by the PT includes\begin{itemize}
    \item \textit{Half-spread}: Distance from current stock price at which player 2 places orders 
    \item \textit{Order side}: Player 2 can choose to place either a single order on the buy or sell side or place on both sides like Player 1 to provide liquidity or hold (and do nothing).
\end{itemize}
Clearly, the action space of player 2 contains that of player 1. This means that player 2 has more capability in choosing its actions than player 1 by the design of their action spaces. Figure \ref{fig:lob} shows an example snapshot of a PT buy order (C) with half-spread $2.5$ cents when the stock price is $10000.5$ cents. 
% \textcolor{red}{Note that a low \textit{Half-spread} corresponds to an order that is more easily executed but results in lower profits for both players. And, orders with a high \textit{Half-spread} give higher profits but have lower chances of being executed.}

The \textbf{reward} for player $i\in\mathcal{N}$ at time $t$ is the change in value of its portfolio from $t-1$ to $t$, given by \begin{align}
    R_i(t)&=\textit{Cash}(t)+\textit{Inventory}(t)\times\textit{Mid-price}(t)\nonumber\\
    &-\textit{Cash}(t-1)-\textit{Inventory}(t-1)\times\textit{Mid-price}(t-1)\nonumber
\end{align}

% \subsection{Outcomes and Social Welfare}
Given the formulation above, we learn policies that maximize discounted cumulative rewards for both players as they interact with each other and the environment for different delays $\delta$. Player outcomes are represented by their profits over the horizon, defined as the difference in portfolio value at $t=H$ to that at $t=0$. Notice that the undiscounted sum of above reward over $H$ precisely gives agent profits. We therefore measure player outcomes by the undiscounted cumulative rewards realized from using learnt policies over $H$. These outcomes are subsequently used to calculate the SWFs in (\ref{eq:swf1}) and (\ref{eq:swf2}). 

% If $\gamma=1$, maximizing the discounted sum of above reward over horizon $H$ corresponds to maximizing the player profits over $H$. We therefore measure player outcomes by the undiscounted cumulative rewards realized from using learnt policies over $H$. These outcomes are used to calculate the SWFs in (\ref{eq:swf1}) and (\ref{eq:swf2}). 

% Given the formulation above, we learn policies that maximize discounted cumulative rewards for both players as they interact with each other and the environment for different delays $\delta$. Aside from analyzing the impact of delay on player strategies and outcomes, we also examine its impact on social welfare. 

\section{EXPERIMENTAL RESULTS}
\subsection{Training}

Our simulated market contains 24 background traders (20 consumer, 4 intelligent), 1 learning MM and 1 learning PT. For every $\delta$, the horizon is a single trading day starting at 9:30am and ending at 4pm. The MM and PT place orders every minute, giving $H=390$ steps per episode. Hence, we vary $\delta$ between $\delta=0$ and $\delta=390$, with $\delta=0$ being the most observable scenario where all states are immediately observable. And, $\delta=390$ being the least observable scenario where a strict sub-part of the states $s_I$ are observable. 
As a caveat, we allow the PT to only place a buy or sell order or hold in the experiments in order to further differentiate between the actions of the MM and PT, although allowing the PT to place on both sides of the market fits the exact formulation.
% Although allowing the PT to place on both sides of the market fits the formulation, PTs in real markets typically trade on a single side. Thus, we allow the PT to only place a buy or sell order or hold in the experiments. 
The background traders come in at random times in the trading day. We use $\gamma=0.9999$ in our experiments since the value of money at the end of the day is nearly the same as that in the beginning. 

\begin{figure}[h!]
    \centering
    \includegraphics[width=\linewidth]{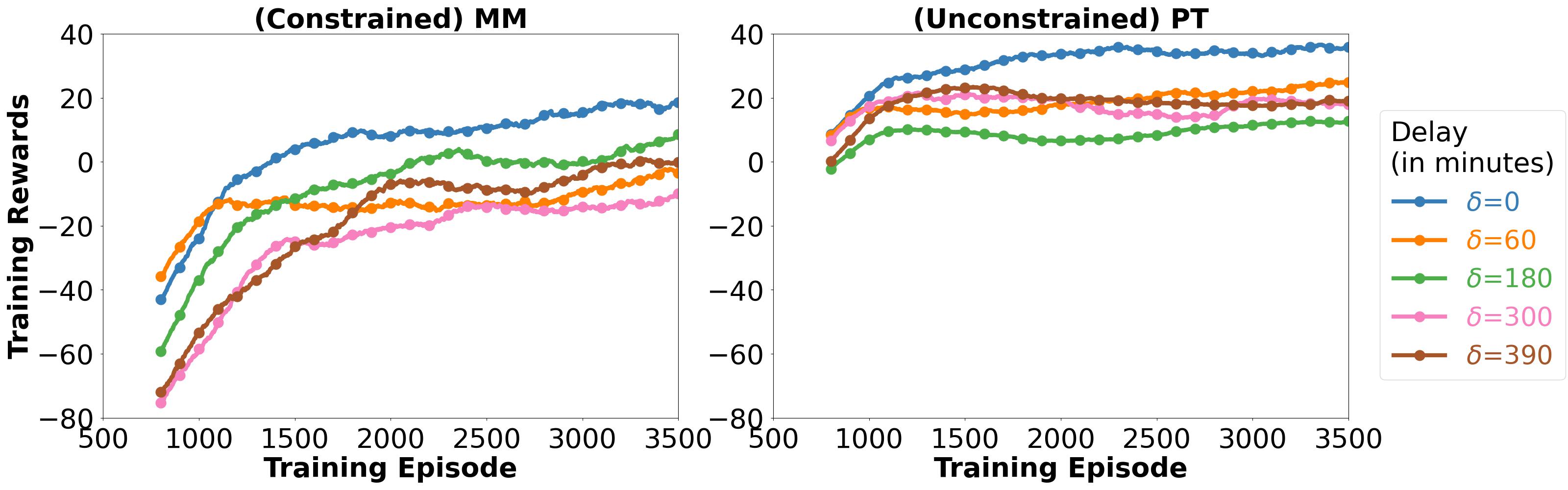}
    \caption{Discounted cumulative rewards while training demonstrating convergence in learning.}
    \label{fig:training_reward}
\end{figure}

Given the formulation in section \ref{subsec:posg_form_markets}, we have continuous states with discrete levels of \textit{Half-spread} and categorical \textit{Order side} options. We use the policy gradient method called Proximal Policy Optimization (PPO) from the RLlib package \shortcite{ppo,rllib} to learn policies for the MM and PT. The states and rewards are normalized to enable efficient exploration in a continuous state, discrete action setup. Figure \ref{fig:training_reward} is a plot of (moving averages of) discounted cumulative rewards of the MM and PT as a function of training episodes. We observe convergence in rewards with training episodes for the values of $\delta$ considered. The difference in scale of rewards between players is because the MM places orders on both sides while the PT typically places on one side or holds, with each order being of equal size.  

\subsection{Impact of Delay on Player Outcomes}

\begin{figure}[h!]
    \centering
    \includegraphics[width=0.95\linewidth]{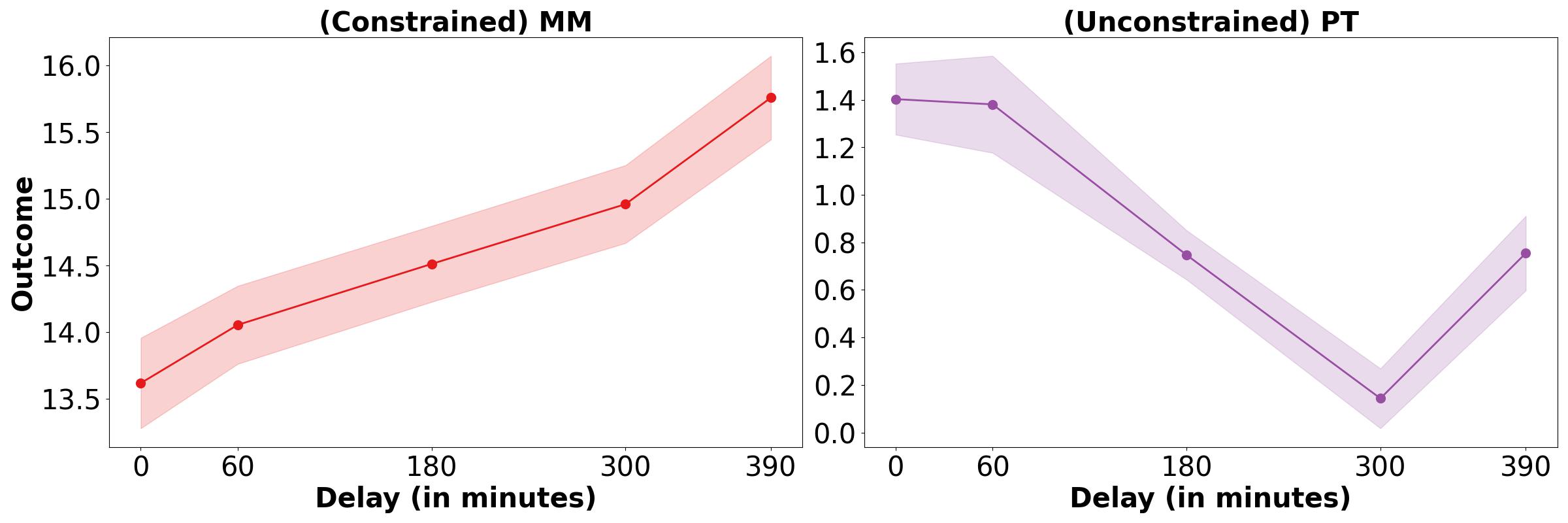}
    \caption{Cumulative rewards measuring player outcomes. With increase in observability (decrease in delay), constrained player outcomes worsen while unconstrained player outcomes improve.}
    \label{fig:pnl}
\end{figure}

The learnt policies of both players are played out in 500 test episodes to collect their realized rewards. Figure \ref{fig:pnl} is a plot of the cumulative rewards of the MM and PT in test episodes (i.e. outcomes) as a function of the delay in observability. The solid line with circular markers represents the mean values, and the shaded region shows the 95\% confidence interval. Interestingly, we observe that the outcomes for the constrained MM improve as delay increases, while those of the PT degrade. Recall that observability increases as the delay $\delta$ decreases (going right to left on the x-axis). This is to say that increased observability empowers the unconstrained player with more information allowing it to improve its outcomes. On the other hand, the constrained player loses out due to inability to act on the new information due to constraints on its actions. Note that the kink at $\delta=300$ arises due to the results at $\delta=390$ varying from the general trend. This is because of the large change in the environment at $\delta=390$ when no delayed states are observable. 

\subsection{Impact of Delay on Learnt Strategies}
\begin{figure}[h!]
    \centering
    \begin{subfigure}{0.5\linewidth}
    \centering
    \includegraphics[width=0.9\linewidth]{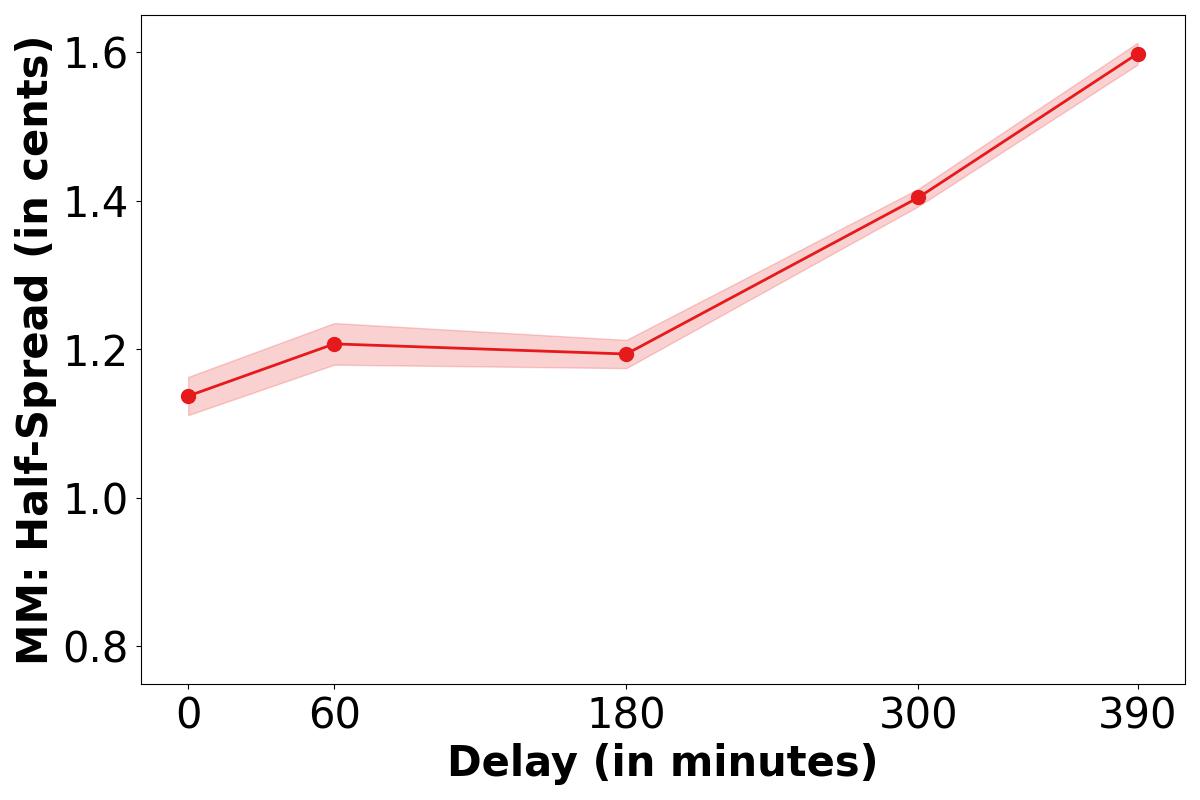}
    \end{subfigure}%
    \begin{subfigure}{0.5\linewidth}
    \centering
    \includegraphics[width=0.9\linewidth]{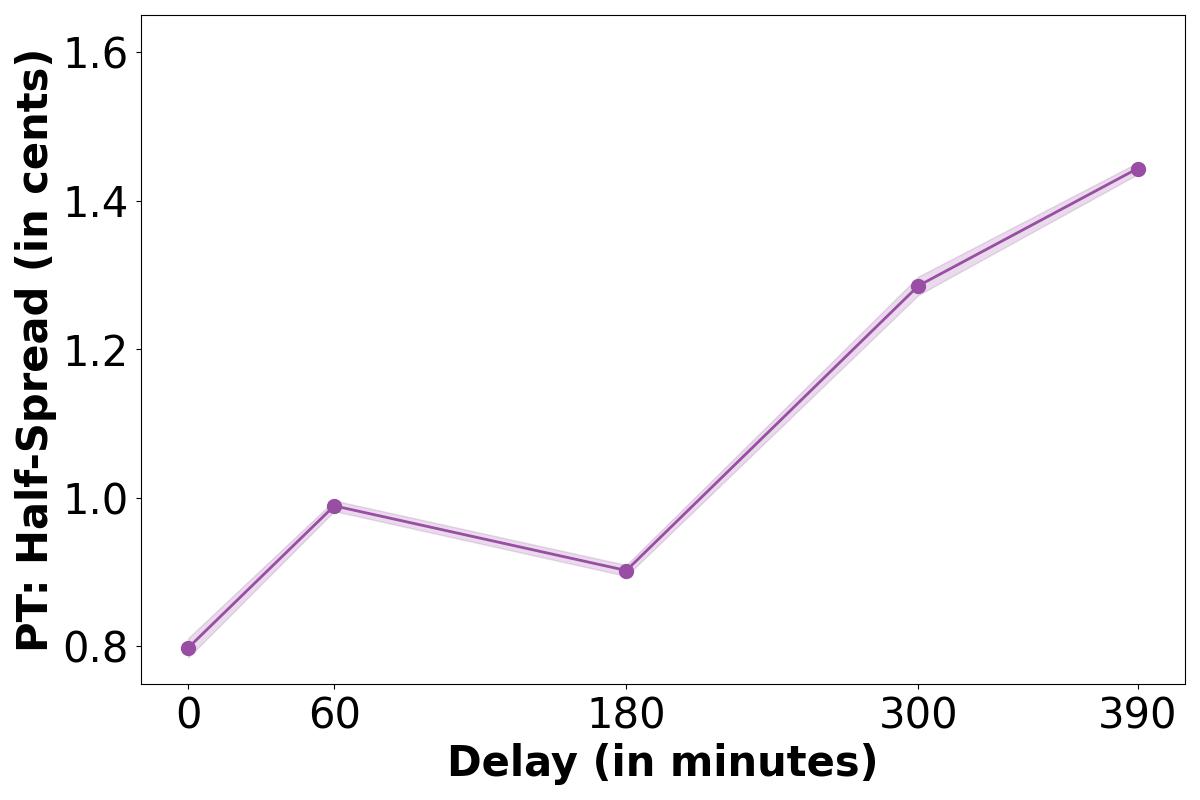}
    \end{subfigure}
    \caption{Learnt strategies of MM and PT: \textit{Half-spread} of orders. See the (near) monotonic trend in strategies with delay.}
    \label{fig:MMPTspread}
\end{figure}

\begin{figure}[h!]
    \centering
    \includegraphics[width=0.45\linewidth]{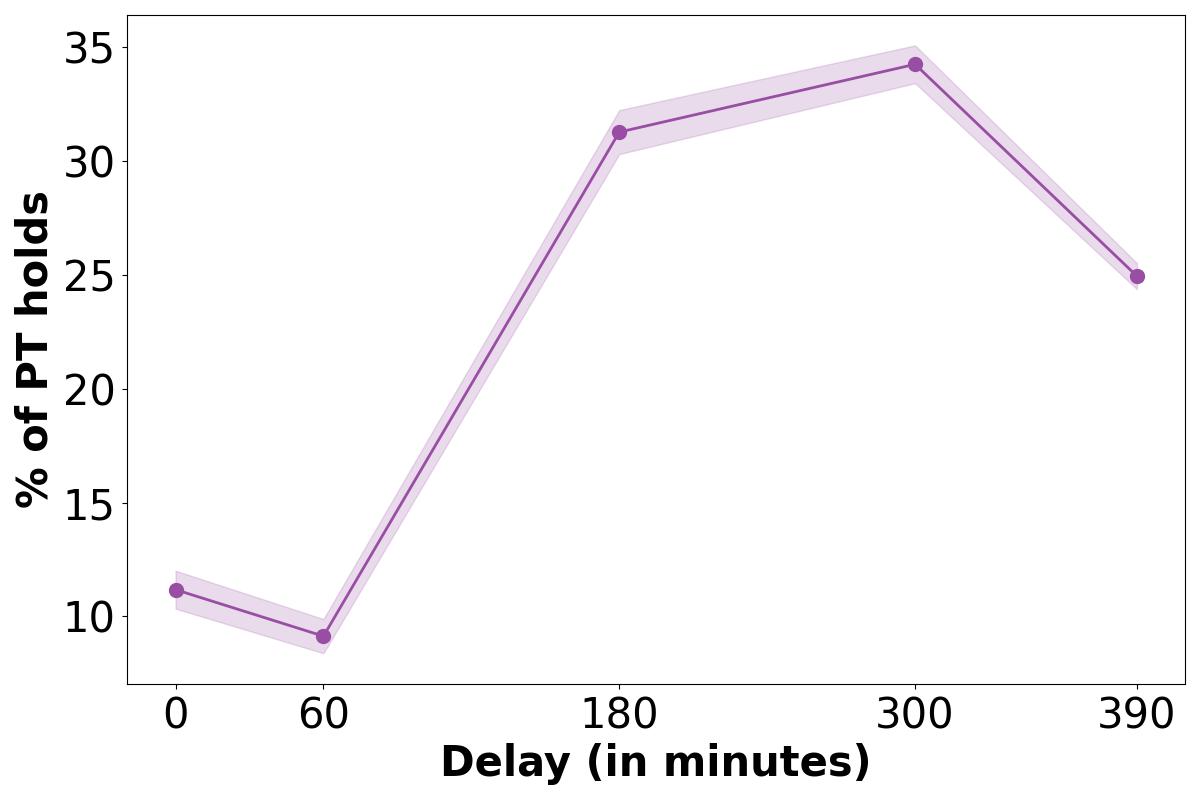}
    % \caption{Learnt strategy of PT: \% of hold decisions. PT learns to trade more frequently (and hold less) at low delays.}
    \caption{Learnt strategy of PT: \% of hold decisions. PT trades more frequently (holds less) at low delays.}
    \label{fig:PTholds}
\end{figure}

% Recall that player actions include \textit{Half-spread} of orders and \textit{Order side}. 
We first plot the average \textit{Half-spread} (average computed across time steps and test episodes per delay) for the MM and PT in Figure \ref{fig:MMPTspread}. We observe a (near) monotonically increasing trend in both strategies as $\delta$ increases, with the PT choosing lower \textit{Half-spread} at every $\delta$ than the MM. 
Orders with lower \textit{Half-spread} have prices closer to the current stock price meaning that they are more competitively priced and have higher chances of being matched.
% than those with less competitive prices.
Therefore, we make the case that increased observability about recent trades allows the PT to place more competitively priced orders than the MM.
The second PT action of \textit{Order side} is categorical and corresponds to multiple directions of placing an order or holding.
% This reasoning about the PT predicting future prices better with more observability is augmented by looking at the PT action of \textit{Order side}. 
% Since \textit{Order side} is categorical and corresponds to multiple directions of placing an order or holding, 
We plot the average \% of hold decisions in a trading day as a function of $\delta$ in Figure \ref{fig:PTholds}. We see that the \% of hold decisions decreases as observability increases. This means the PT places buy/sell orders more frequently at low $\delta$, and prefers to hold more at high $\delta$. Thus, observability enables the PT to trade more frequently. 

% predict for future prices better, enabling it to take calculated risks and place at lower \textit{Half-spreads}.
% Therefore, we make the case that the PT starts placing closer to the stock price as observability increases (right to left), with the MM following suit.

% Also, noting that the PT chooses lower \textit{Half-spread} at every $\delta$ than the MM, we make the case that the PT starts placing closer to the stock price as observability increases (right to left), with the MM following suit. 
% Thus, increased observability about recent trades allows the PT to predict for future prices better, enabling it to take calculated risks and place at lower \textit{Half-spreads}. These competitively priced orders of the PT urge the MM to also reduce its \textit{Half-spread} to increase its chances of execution. At the same time, since the MM places on both sides of the market, this can be disadvantageous when the price moves since one side of its orders are now exposed. Effectively to say that the PT profits go up and MM profits go down with increase in observability. 

\begin{figure}[h!]
    \centering
    \begin{subfigure}{0.5\linewidth}
    \centering
    \includegraphics[width=\linewidth]{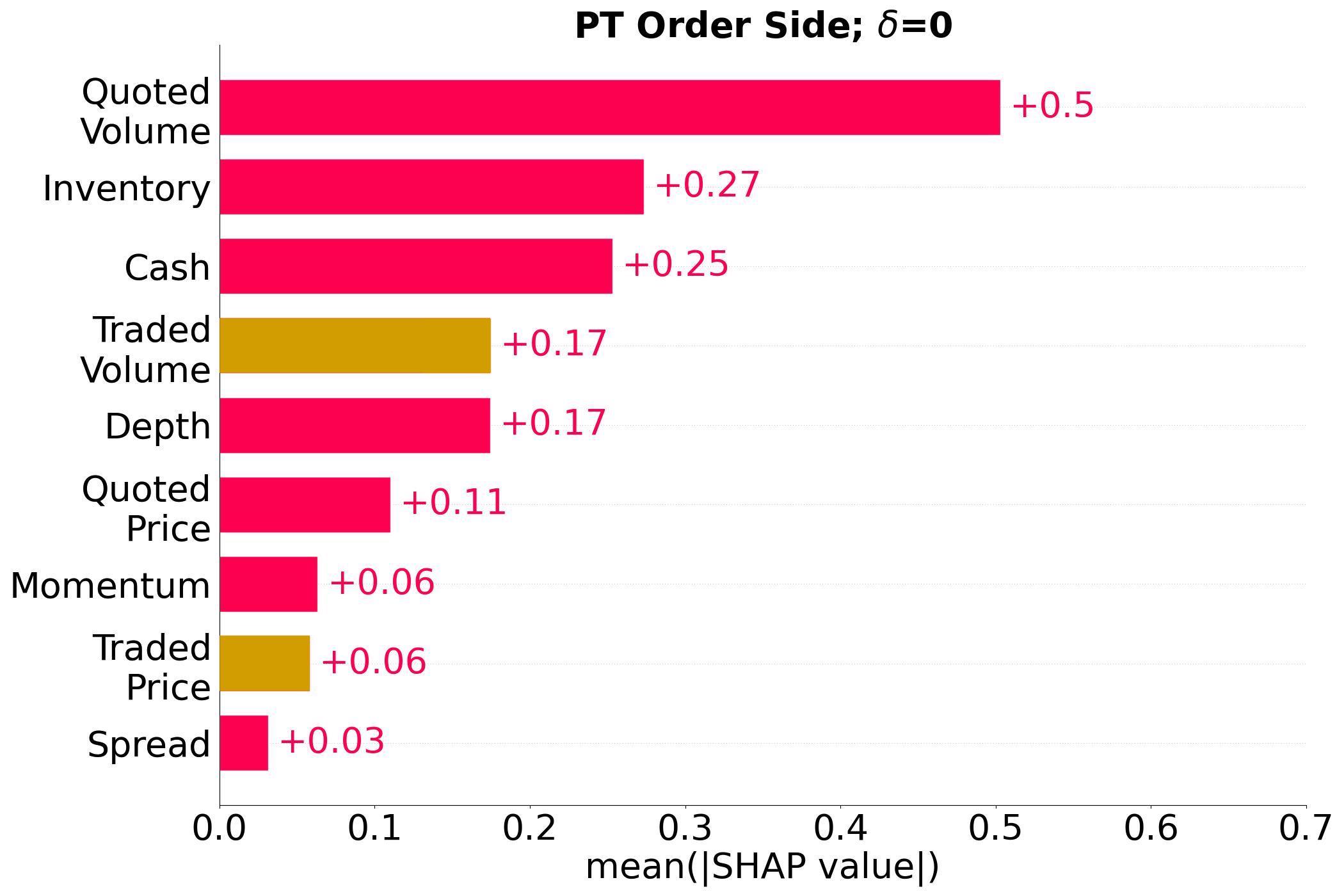}
    % \caption{}
    % \label{fig:pt_a1_d0}
    \end{subfigure}%
    \begin{subfigure}{0.5\linewidth}
    \includegraphics[width=\linewidth]{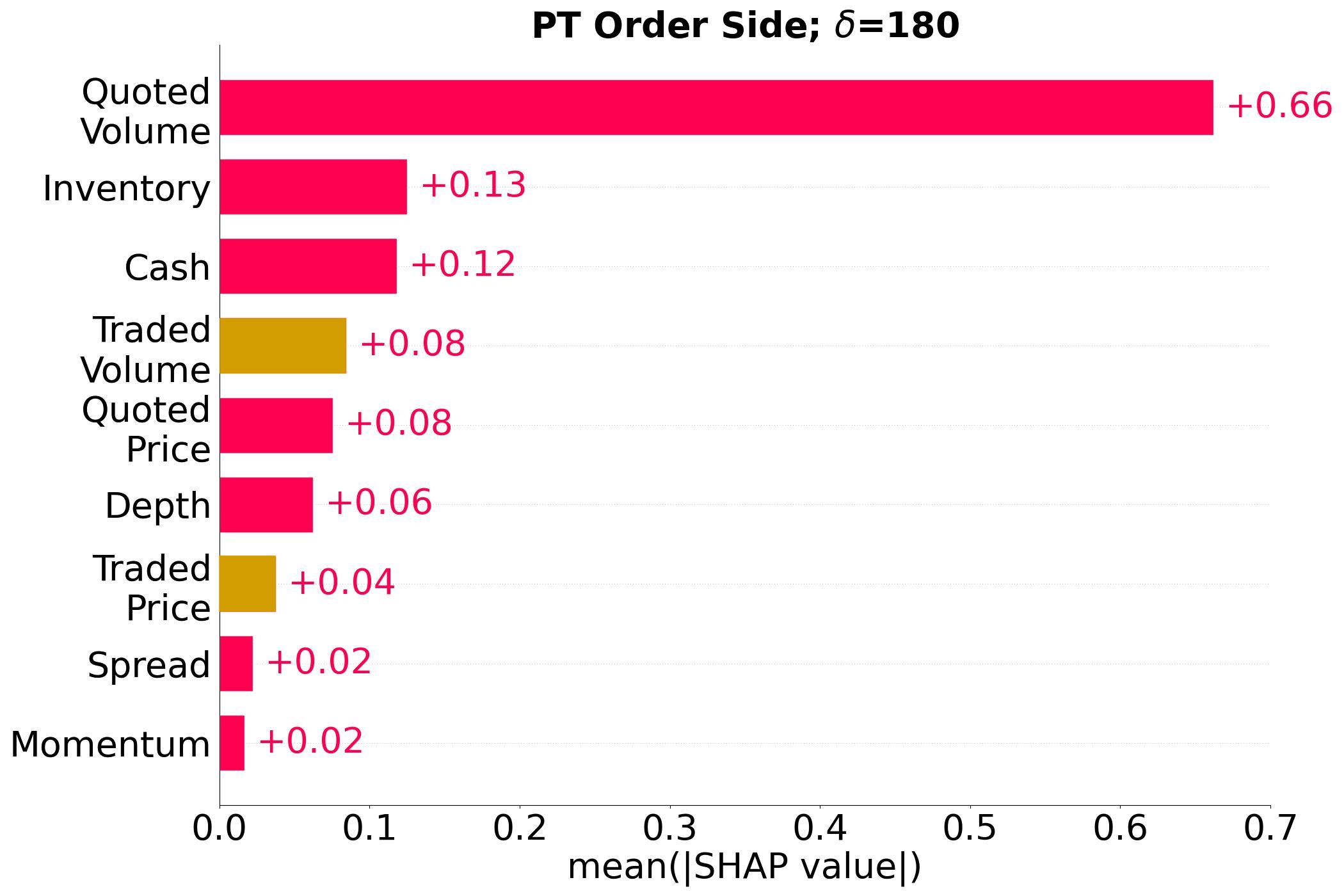}
    % \caption{}
    % \label{fig:pt_a1_d180}
    \end{subfigure}
    \caption{Impact of features on PT action of \textit{Order side} for $\delta=0$ (left) and $\delta=180$ (right). See the reduction in importance for delayed features of `\textcolor{olive}{Traded Volume}' and `\textcolor{olive}{Traded Price}' with increase in $\delta$.} 
    \label{fig:pt_a1}
\end{figure}

To investigate which observation features are most impactful towards PT decisions, we use the explainability tool called SHAP (for SHapley Additive exPlanations). We examine the PPO policy network that takes in observations to give out PT actions. SHAP uses cooperative game theory to decompose the network output locally into a sum of effects attributed to each input feature \cite{shap}. SHAP values are scores measuring the importance of observation features towards the action prediction. For illustration, we use SHAP for the PT action of \textit{Order side} to calculate global feature importances for market observations (section \ref{subsec:posg_form_markets}) in Figure \ref{fig:pt_a1}. Note that delayed features of traded volume and traded price are colored in \textcolor{olive}{olive}. These global importances are calculated as the average of absolute SHAP values across a dataset comprising (observation, action) pairs. 

The left plot in Figure \ref{fig:pt_a1} shows feature importances when $\delta=0$, where quoted volume is the most impactful feature with traded volume coming in fourth. The right plot in Figure \ref{fig:pt_a1} shows feature importances when $\delta=180$, where the importance scores for traded volume and traded price decrease from their values in the left plot. Thus, the delayed features lose their importance in defining actions as the delay is increased. 
% This is in line with what we expect as the delay is increased i.e., the PT starts using the traded volumes feature less as the delay in that information is increased. 
We also see that the importance of quoted volume increases as the delay in traded volume is increased. This goes to say that the PT uses quoted volume as a substitute for traded volume when the latter is delayed.

\subsection{Impact of Delay on Social Welfare}
\begin{figure}[h!]
    \centering
    \includegraphics[width=0.87\linewidth]{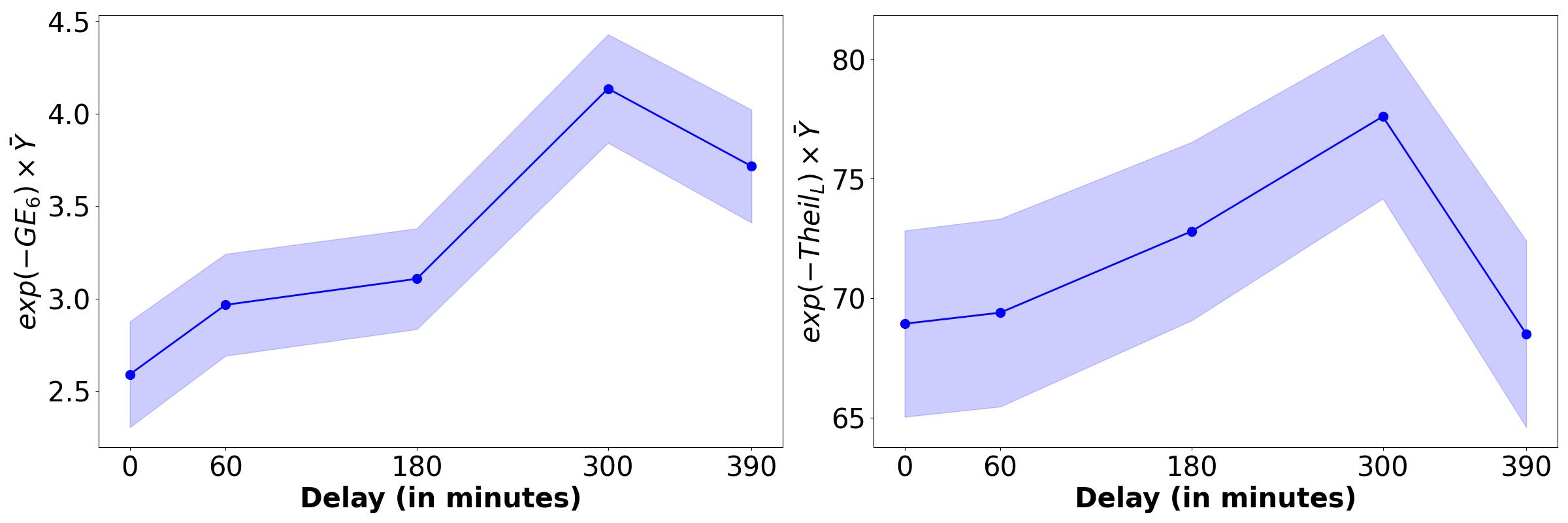}
    \caption{Social Welfare as measured by (\ref{eq:swf1}) in the left plot and (\ref{eq:swf2}) in the right plot. See that social welfare is maximized at intermediate values of delay in both cases.}
    \label{fig:SWF}
\end{figure}

% Figure \ref{fig:SWF} shows the SWF in (\ref{eq:swf1}) on the left with $\kappa=6$ and that in (\ref{eq:swf2}) on the right, as functions of $\delta$. 
Figure \ref{fig:SWF} is a plot of  SWFs computed using agent outcomes in test episodes as a function of $\delta$. The left plot uses (\ref{eq:swf1}) with $\kappa=6$ while the right plot uses (\ref{eq:swf2}). 
We observe both SWFs increase with $\delta$ up to $\delta=300$ after which they fall down. Thus, there lies an intermediate delay at which social welfare is maximized. 

\subsection{Discussion}
In our setup, $\delta=390$ is a fully opaque regime when no delayed states are observable, with $\delta=0$ being a fully transparent regime when all states are immediately observable. 
With increase in transparency, both agents were seen to place more competitively priced orders with lower \textit{Half-spread} that in turn reduces the market spread. 
Social welfare was found to be highest at intermediate $\delta$ that correspond to a partial transparency regime. This can provide an indication to policymakers on optimal transparency regimes that may not be intuitive by looking at outcomes of agents alone. Although our experiments were performed in a generic exchange market simulator that is not calibrated to a specific financial market, similar findings have been observed with the introduction of transparency in real markets as shown in Table \ref{tab:real_trends}. 

\begin{table}[h!]
    \centering
    \caption{Trends that have been observed in real markets with increase in transparency. 
    Our findings in simulated markets are in line with those observed in real markets.}
    \label{tab:real_trends}
    \begin{tabular}{|c|c|c|c|c|}\hline
        Metric & Definition & Market & Trend in real data & Trend in our work \\\hline
        \multirow{2}{3.7em}{\centering Price dispersion} & Volume weighted difference & \multirow{2}{5.6em}{\centering Interest Rate Swaps} & \multirow{2}{7.8em}{\centering Fell by 12-19\% \shortcite{benos2020centralized}} & \multirow{2}{5.8em}{\centering Market spread reduced}\\
        % & & & & \\
        % & & & & \\
        & between mid-price, traded price & & & \\\hline
        \multirow{2}{3.7em}{\centering Bid-ask spread} & Difference between best buy, & \multirow{2}{5.6em}{\centering US Corporate Bonds} & \multirow{2}{7.8em}{\centering Reduced \shortcite{mizrach2015analysis}} & \multirow{2}{5.8em}{\centering Reduced}\\
        % & & & & \\
        & best sell prices & & & \\\hline
    \end{tabular}
\end{table}

\section{CONCLUSION}

% For example, this would mean that if the PT predicts the price to increase in the future, it would place a buy order close to the current stock price. This compels the MM to also place buy and sell orders close to the current stock price to ensure execution and hence non-zero profits. 
%This behavior of predicting for price moves is also seen when the learnt strategies are played on sample test episodes shown in Figure. 

We consider the problem of defining and evaluating the impact of transparency in multi-agent systems comprising adaptive agents using simulations. By defining transparency (or lack thereof) as delay in observability for agents, we propose a multi-agent reinforcement learning framework to evaluate the effects of varying observability on agent strategies and social welfare. We specifically look at the interplay between constrained and unconstrained agents. The framework is illustrated with experiments in simulated financial markets comprising constrained and unconstrained traders. We observe that increasing observability improves outcomes for unconstrained agents albeit with degrading outcomes for the constrained agents. %We also identify an optimal delay in observability at which social welfare is maximized.  
We also experimentally demonstrate that social welfare is maximized at intermediate values of delay.

\section*{ACKNOWLEDGMENTS}
This paper was prepared for informational purposes by the Artificial Intelligence Research group of JPMorgan Chase \& Co and its affiliates (``J.P. Morgan'') and is not a product of the Research Department of J.P. Morgan. J.P. Morgan makes no representation and warranty whatsoever and disclaims all liability, for the completeness, accuracy or reliability of the information contained herein. This document is not intended as investment research or investment advice, or a recommendation, offer or solicitation for the purchase or sale of any security, financial instrument, financial product or service, or to be used in any way for evaluating the merits of participating in any transaction, and shall not constitute a solicitation under any jurisdiction or to any person, if such solicitation under such jurisdiction or to such person would be unlawful.   

% Reducing font size (to 9pt) for References & Author Biagraphies
\footnotesize

% Please don't exchange the bibliographystyle style
\bibliographystyle{wsc}

% AUTHOR: Include your bib file here
\bibliography{wsc23paper}

\section*{AUTHOR BIOGRAPHIES}

\noindent {\bf KSHAMA DWARAKANATH} is a Research Scientist at J.P. Morgan AI Research working on using reinforcement learning to design and learn trading agents with diverse objectives in simulated multi-agent markets. Her interests lie in the fields of reinforcement learning, multi-agent simulations and mechanism design. Her email address is \email{kshama.dwarakanath@jpmorgan.com}.\\

\noindent {\bf SVITLANA VYETRENKO} is an Executive Director at J.P. Morgan AI Research leading a team focusing on generative time series models, multi-agent simulations and reinforcement learning. Her email address is 
\email{svitlana.s.vyetrenko@jpmchase.com}.\\

\noindent {\bf TOKS OYEBODE} is an Executive Director in Regulatory Affairs for J.P. Morgan’s Corporate \& Investment Bank (CIB). He leads CIB's policy engagement on regulatory developments impacting derivatives and fixed income market structure. 
Toks was a Technical Specialist at the UK’s Prudential Regulation Authority. His email address is \email{toks.oyebode@jpmorgan.com}.\\
% Prior to joining J.P. Morgan, Toks was a Technical Specialist at the UK’s Prudential Regulation Authority. His email address is \email{toks.oyebode@jpmorgan.com}.\\
% Prior to joining J.P. Morgan, Toks was a Technical Specialist at the UK’s Prudential Regulation Authority focusing on banks’ liquidity risk management. 
% He is a CFA Charterholder, and his email address is \email{toks.oyebode@jpmorgan.com}.\\

\noindent {\bf TUCKER BALCH} is a Research Director at J.P. Morgan AI Research and a professor of Interactive Computing at Georgia Tech (on leave). He is interested in problems concerning multi-agent social behavior in domains ranging from financial markets to tracking and modeling the behavior of ants, honeybees and monkeys. 
% He co-founded Lucena Research, an investment software firm that applies Machine Learning and Big Data approaches to investment problems. 
His email address is \email{tucker.balch@jpmchase.com}.

% \section{To Do}
% \begin{enumerate}
%     \item Figure 6 - font size to be Arial (8pt + )
%     \item Page limit
% \end{enumerate}

\end{document}